\documentclass[11pt,a4paper]{article}
\pdfoutput=1
\usepackage{jcappub}
\usepackage{amsmath}
\usepackage{graphicx}
\usepackage{caption}
\usepackage{subcaption}
\usepackage{placeins}
\usepackage{float}
\usepackage[english]{babel}
\usepackage{epstopdf}

\title{Power Spectra beyond the Slow Roll Approximation in Theories with Non-Canonical Kinetic Terms}
\author{Carsten van de Bruck}
\author{and Mathew Robinson}
\affiliation{Consortium for Fundamental Physics, School of Mathematics and Statistics, University of Sheffield,
Hounsfield Road, Sheffeld, S3 7RH, United Kingdom}
\emailAdd{C.vandeBruck@sheffield.ac.uk}
\emailAdd{app11mrr@sheffield.ac.uk}
\abstract{We derive analytical expressions for the power spectra at the end of inflation in theories with two inflaton fields and non-canonical kinetic terms. We find that going beyond the slow--roll approximation is necessary and that the nature of the non-canonical terms have an important impact on the final power spectra at the end of inflation. We study five models numerically and find excellent agreement with our analytical results. Our results emphasise the fact that going beyond the slow--roll approximation is important in times of high-precision data coming from cosmological observations.}
\arxivnumber{}
\begin{document}
\maketitle%
\section{Introduction}

Whilst inflation has enjoyed multiple successes in explaining the observed large scale structure, formed by density perturbations in the early universe, there is still much work to be done in narrowing down the specifics regarding the types of models still viable in this age of precision cosmology --- especially in light of the current data coming from Planck \cite{Planck06,Planck13} and BICEP2 \cite{BICEP2} on the properties of the Cosmic Microwave Background radiation (CMB). Early models involved only a single scalar field, the inflaton \cite{Guth81,Linde82} rolling down to it's minimum to drive inflation, but more recently multi-field models have come to the fore motivated by theoretical considerations in high energy physics \cite{Kofman87,SilkTurner,Wands96,Staro1,Shiu2011,White,Kaiser1,Kaiser2,Kallosh2013,Kaiser3,Kaiser4,Riberio}. There is no clear reason that inflation should be driven by only a single field so it is natural to study this class of models equally --- especially when they lead to a much richer and more diverse set of predictions. In addition to using multiple scalar fields, these higher energy theories suggest that the inclusion of non-standard kinetic terms is a natural progression from the standard models --- and that's what we shall look at here. 

Studying the evolution of perturbations (and power spectra) of single field models is a relatively simple task in that they become frozen in as the mode of interest exits the horizon. This is not, however, the case with multi-field models in which the added degree of freedom allows for perturbations in two directions --- those along the direction of the background trajectory (curvature perturbations) and those orthogonal (isocurvature perturbations) --- which are not mutually independent as inflation proceeds. Isocurvature perturbations describe the relative perturbations between the fields present and it has already been shown (see e.g. \cite{GarcWands,Starob95,Wands2}) that curvature perturbations can be sourced by isocurvature perturbations long after horizon exit, thereby adding an additional layer of complexity to finding the values of the power spectra come the end of inflation. This sourcing of perturbations is strongest when there are sharp turns in field space and we expect that the coupling between the fields will both lead to additional curvature in the background trajectory in a number of cases along with another independent source term \cite{DiMarco03}, in turn leading to additional curvature \cite{Lalak07,DiMarco05}. It is therefore important to be able to track the perturbations and their mixing throughout inflation to ascertain a faithful estimate of their final values in order to compare to observational data. 

A significant amount of work has already been done on this \cite{Langlois99,Bartolo01,Byrnes06,Cremonini10,Cremonini11, Hu11}  both using the $\delta N$ formalism \cite{Starob85,Sasaki96,Lee05,Sugiyama12} and the transfer matrix method \cite{Lalak07,Davis12} where it has been noted that in some circumstances the $\delta N$ formalism is unsuitable and so we take a closer look at the transfer matrix method and attempt to generalize the work from \cite{DiMarco03,DiMarco05,Lalak07,Davis12} to include terms second order in slow-roll parameters in both canonical and non-canonical cases. Previous work has largely focused on higher order approximations in purely canonical cases or first order approximations in non-canonical cases so the logical next step is to bring these two situations together. We find that due to the relative sizes of some of the non-canonical slow roll parameters (calculated from the background trajectories), introduced later, this step is necessary and well motivated as non-canonical terms can dominate over their standard counterparts. 

In this paper we examine the effects of these non-canonical terms at second order in two distinct regimes --- the early time/horizon crossing regime along with the super-horizon evolution regime. Splitting the evolution up in this way is necessary because of the very different behaviours during these times. The early evolution is dominated by an explicit time dependence with (generally) very small slow roll paramaters. After this, we expect the explicit time dependence to vanish and the evolution to be dominated purely by the growing slow roll terms. The final results given here shall combine both of these approaches, with the results normalised consistently for ease of comparison with earlier work.

The paper is organised as follows. The next section gives details of the background and the form of the models to be looked at along with an introduction to the slow roll parameters used later on. It then details the early times regime, calculating the resulting power spectra approximations before continuing on to the super-Hubble regime and finding analytic approximations there too. The calculation presented in this section follows  \cite{Lalak07} closely, but going to second order in the slow-roll parameter. Section 3 introduces the numerical procedure along with the inflationary potentials we shall be looking at. The fourth section looks at the results of the numerics and compares them to the first and second order analytic approximations found earlier in a number of cases. Finally, we give our conclusions based on the results of section 4. In the appendix, the reader can find some formulae which are needed in the main text. 

\section{Theory and analytical results}

The model considered in this paper includes two scalar fields --- one of which has a non-canonical kinetic term --- and is described by the following action,
\begin{equation}
S = \int{d^4x\sqrt{-g}\left[\frac{M_{\rm P}^2}{2}R-\frac{1}{2}(\partial_\mu\phi)(\partial^\mu\phi)-\frac{e^{2b(\phi)}}{2}(\partial_\mu\chi)(\partial^\mu\chi)-V(\phi,\chi) \right]}.
\end{equation}
In the rest of the paper we set $M_{\rm P} = 1$. The expressions we derive in this section are valid for a general function $b(\phi)$ and the background motion of the fields is determined by
\begin{eqnarray}
\ddot{\phi}+3H\dot{\phi}+V_\phi &=& b_\phi e^{2b}\dot{\chi}^2, \\
\ddot{\chi}+(3H+2b_\phi \dot{\phi})\dot{\chi}+e^{-2b}V_\chi &=& 0, 
\end{eqnarray}
in which both the dot and $H$ are given with respect to cosmic time, $t$. Finally, the Friedmann equations now take the form
\begin{eqnarray}
\dot{H} &=& -\frac{1}{2M_P^2}\left[\dot{\phi}^2+e^{2b}\dot{\chi}^2 \right]\quad \text{and}\\
H^2 &=& \frac{1}{3M_P^2}\left[\frac{\dot{\phi}^2}{2} + \frac{e^{2b}}{2}\dot{\chi}^2 + V \right].
\end{eqnarray}
In order to study perturbations at linear order in this theory, we work in the longitudinal gauge in which the perturbed metric is given by
\begin{equation}
ds^2 = -(1+2\Phi)dt^2 + a^2(1-2\Phi)dx^2
\end{equation}
and the scalar fields broken down into their background and perturbed components
\begin{eqnarray}
\phi(t,x) &=& \phi(t)+\delta\phi(t,x)\quad \text{and} \\
\chi(t,x) &=& \chi(t)+\delta\chi(t,x).
\end{eqnarray}
From these we can find the perturbed Klein-Gordon equations and associated energy and momentum constraints via the Einstein equations. The next step is to perform an instantaneous rotation in field space \cite{Gordon01} on $\phi$ and $\chi$:
\begin{eqnarray}
\delta\sigma &\equiv& \cos\theta \delta\phi+\sin\theta e^b\delta\chi,\\
\delta s &\equiv& -\sin\theta\delta\phi + \cos\theta e^b \delta\chi,
\end{eqnarray}
where we have defined
\begin{equation}
\cos\theta= \frac{\dot{\phi}}{\dot{\sigma}}, \quad\quad\quad\quad \sin\theta = \frac{\dot{\chi}e^b}{\dot{\sigma}}, \quad\quad\text{and}\quad\quad \dot{\sigma} = \sqrt{\dot{\phi}^2+e^{2b}\dot{\chi}^2}.
\end{equation}
In this new basis, we can see that $\delta\sigma$ equates to a perturbation along the same direction as the background trajectory at that point, $\delta s$ equates to a perturbation orthogonal to the background trajectory and $\dot{\theta}$ describes the amount of curvature.
At this point it is useful to work with the gauge invariant Mukhanov-Sasaki variables \cite{Sasaki86,Mukhanov88}, defined by
\begin{eqnarray}
Q_\sigma &\equiv& \delta\sigma - \frac{\dot{\sigma}}{H}\Phi,
\end{eqnarray}
The isocurvature perturbation $\delta s$ is automatically gauge invariant by the Stewart--Walker lemma. For completeness, in this basis the background equations become 
\begin{eqnarray}
\ddot{\sigma} + 3H\dot{\sigma} + V_\sigma &=& 0 \quad \text{and} \\
\dot{\theta} + \frac{V_s}{\dot{\sigma}} + b_\phi \dot{\sigma}\sin(\theta) &=& 0,
\end{eqnarray}
whilst the perturbation equations can be written as
\begin{eqnarray}\label{exacteq}
\left(\ddot{Q_\sigma} \atop \ddot{\delta s}\right) + 
\left(\begin{array}{cc}3H & \frac{2V_{,s}}{\dot{\sigma}}\\-\frac{2V_{,s}}{\dot{\sigma}} & 3H\end{array}\right)
\left(\dot{Q_\sigma} \atop \dot{\delta s}\right) + \left[\frac{k^2}{a^2}\textbf{1} +
\left(\begin{array}{cc}C_{\sigma\sigma} & C_{\sigma s} \\C_{s\sigma} & C_{ss}\end{array}\right)\right]
\left(Q_\sigma \atop \delta s\right) = 0.
\end{eqnarray}
The coefficients, $C_{AB}$ are given below:

\begin{eqnarray}  \label{exactC}
C_{\sigma\sigma} &=& V_{\sigma\sigma}-\left(\frac{V_s}{\dot{\sigma}}\right)^2+2\frac{\dot{\sigma}V_\sigma}{HM^2} + 3\frac{\dot{\sigma}^2}{M^2}-\frac{\dot{\sigma}^4}{2M^4H^2} - b_\phi \left(s^2_\theta c_\theta V_\sigma + \left(c^2_\theta + 1 \right)s_\theta V_s \right),\label{csigmasigma}\\
C_{\sigma s} &=& 6H\frac{V_s}{\dot{\sigma}} + 2\frac{V_\sigma V_s}{\dot{\sigma}^2}+2V_{\sigma s}+\frac{\dot{\sigma}V_s}{M^2H}+2b_\phi\left(s^3_\theta V_\sigma - c^3_\theta V_s\right),\\
C_{s \sigma} &=& -6H\frac{V_s}{\dot{\sigma}}-2\frac{V_\sigma V_s}{\dot{\sigma}^2}+\frac{\dot{\sigma}V_s}{M^2H}, \label{cssigma}\\
C_{ss} &=& V_{ss} - \left(\frac{V_s}{\dot{\sigma}}\right)^2 + b_\phi\left(1+s^2_\theta\right)c_\theta V_\sigma + b_\phi c^2_\theta s_\theta V_s - \dot{\sigma}^2\left(b_{\phi\phi}+b^2_\phi\right),
\end{eqnarray}
where $s_\theta = \sin(\theta)$ and $c_\theta = \cos(\theta)$.

\subsection{Slow-Roll-Inflation}
From the background equations we define the slow roll parameters in the usual way:
\begin{eqnarray}
\epsilon = -\frac{\dot{H}}{H^2}   \quad \text{and}\\
\eta_{AB} = \frac{V_{AB}}{3H^2}
\end{eqnarray}
However, in this paper we shall need expressions for the slow roll parameters at 2nd order as we no longer assume they remain constant throughout. We find similar expressions to those found in \cite{Davis12} with the additional terms related to the non-canonical components. Following the definition of $\xi$ used in \cite{Lalak07}, we write 
\begin{eqnarray}
\xi_1 &=& \sqrt{2\epsilon}b_\phi \quad \text{and}\\
\xi_2 &=& 2\epsilon b_{\phi\phi},
\end{eqnarray}
which shall be treated as first and second order slow roll parameters respectively. The time derivatives of the slow roll parameters then become
\begin{eqnarray}
\dot{\eta_{\sigma\sigma}} &=& 2H\epsilon\eta_{\sigma\sigma} - 2H\eta^{2}_{\sigma s}-2H\eta_{\sigma\sigma}\xi_1 s^2_\theta c_\theta -4H\eta_{\sigma s}\xi_1 s_\theta c_\theta^2 - H\alpha_{\sigma\sigma\sigma},\\
\dot{\eta_{\sigma s}} &=& 2H\epsilon\eta_{\sigma s} +H\eta_{\sigma s}\eta_{\sigma \sigma}-H\eta_{\sigma s}\eta_{ss} -2H\eta_{ss}\xi_1 s_\theta c_\theta^2 -H\eta_{\sigma s}\xi_1 c_\theta - H\alpha_{\sigma\sigma s}, \\
\dot{\eta_{ss}}&=& 2H\epsilon\eta_{ss}+2H\eta_{\sigma s}^2-2Hc^3_\theta\xi_1\eta_{ss} - H\alpha_{\sigma ss} \quad \text{and}\\
\dot{\xi_1} &=& 2H\epsilon\xi_1 - H\xi_1\eta_{\sigma\sigma}- H\xi_1^2 s^2_\theta c_\theta + H\xi_2c_\theta \label{xi1dot},
\end{eqnarray}
where 
\begin{equation}
\alpha_{IJK}\equiv \frac{V_\sigma V_{IJK}}{V^2}.
\end{equation}
We can now begin work on the perturbation equations by expressing the $C_{AB}$ coefficients in terms of the slow roll parameters up to second order. It is in this form that we shall use them from now on, rather than the exact equations given earlier, in equation (\ref{exactC}). They are given by 
\begin{equation}
\begin{split}
C_{\sigma\sigma} = 3H^2\left[\eta_{\sigma\sigma}-2\epsilon+\xi_1 s^2_\theta c_\theta - \frac{\eta^2_{\sigma s}}{3}-2\epsilon^2 +\frac{4\epsilon\eta_{\sigma\sigma}}{3}+\frac{\xi_1\eta_{\sigma s}}{3}(s_\theta-3s_\theta c_\theta^2)\quad\quad\quad\right.\\
\left. \quad\quad +\frac{5\epsilon\xi_1 s^2_\theta c_\theta}{3} - \frac{\xi_1\eta_{\sigma\sigma}s^2_\theta c_\theta}{3}+\frac{\xi_1^2s_\theta^4 c_\theta^2}{3}  \right],
\end{split}
\end{equation}
\begin{equation}
C_{\sigma s} = 3H^2\left[2\eta_{\sigma s} - 2\xi_1 s^3_\theta +\frac{2\eta_{\sigma\sigma}\eta_{\sigma s}}{3}-\frac{2\epsilon\xi_1 s^3_\theta}{3}+\frac{2\xi_1^2 c^3_\theta s^3_\theta}{3} + \frac{2\eta_{\sigma s}\xi_1 c_\theta(s^2_\theta - c^2_\theta)}{3}\;\right],
\end{equation}
\begin{equation}
C_{s \sigma} = 3H^2 \left[\frac{4\epsilon\eta_{\sigma s}}{3} - \frac{2\eta_{\sigma\sigma}\eta_{\sigma s}}{3}+\frac{2\eta_{\sigma\sigma}\xi_1 s^3_\theta}{3}-\frac{4\epsilon\xi_1 s^3_\theta}{3} -\frac{2\eta_{\sigma s}\xi_1 s^2_\theta c_\theta}{3}+ \frac{2\xi_1^2 s^5_\theta c_\theta}{3}\;\; \right] , \\
\end{equation}
\begin{equation}
\begin{split}
C_{ss} = 3H^2\left[\eta_{ss}-\xi_1(1+s^2_\theta)c_\theta -\frac{\eta_{\sigma s}^2}{3}+\frac{\xi_1^2 c^2_\theta(s^4_\theta-1)}{3}+\frac{\eta_{\sigma s}\xi_1 s_\theta(1+s_\theta^2)}{3}\quad\quad\quad\right.\\
\left. \quad\quad +\frac{\eta_{\sigma\sigma}\xi_1 c_\theta(1+s^2_\theta)}{3} -\frac{\epsilon\xi_1 c_\theta(1+s^2_\theta)}{3} - \frac{\xi_2}{3}\right],
\end{split}
\end{equation}
In order to track the perturbations and calculate the resulting power spectra, we split the calculation into two regimes: the horizon crossing regime which is applicable as the mode of interest exits the horizon and the subsequent evolution --- from a few e-folds after horizon crossing through until the end of inflation. 

\subsection{Horizon Crossing}

Using the substitutions $u_\sigma = aQ_\sigma$ and $u_s = a\delta s$ to rewrite Eq.~(\ref{exacteq}) in conformal time we find 
\begin{eqnarray}
\left[\left( \frac{d^2}{d\tau^2}+k^2-\frac{a''}{a} \right)\mathbf{1}+2\mathbf{E}\frac{1}{\tau}\frac{d}{d\tau}+\mathbf{M}\frac{1}{\tau^2}\right]\left(u_\sigma \atop u_s \right) = 0, \label{matrixeq}
\end{eqnarray}
where
\begin{eqnarray}
\mathbf{E} &=& \left(\begin{array}{cc}0 & \frac{aV_{s}}{\dot{\sigma}}\\-\frac{aV_{s}}{\dot{\sigma}} & 0\end{array}\right)\\
\mathbf{M} &=& \left(\begin{array}{cc}a^2C_{\sigma\sigma} & a^2C_{\sigma s} -\frac{2a'V_{s}}{\dot{\sigma}} \\a^2C_{s \sigma}+\frac{2a'V_{s}}{\dot{\sigma}} & a^2C_{ss}\end{array}\right) \label{meq}
\end{eqnarray}
and the $\frac{a''}{a}$ term which can be expanded in terms of the slow--roll parameters to second order as
\begin{eqnarray}
\frac{a''}{a} = \frac{1}{\tau^2}\left[ 2+3\epsilon+20\epsilon^2-8\epsilon\eta_{\sigma\sigma}-8\epsilon\xi_1 s^2_\theta c_\theta \right].\label{adoubledash}
\end{eqnarray}
For the moment we will be neglecting the $\frac{a''}{a}$-term and with the help of the rotation given by
\begin{eqnarray}
\mathbf{R} = \left(\begin{array}{cc}\cos\Theta & -\sin\Theta\\ \sin\Theta & \cos\Theta\end{array}\right),
\end{eqnarray}
it is possible to again rewrite Eq.(\ref{matrixeq}), which is of the form $u'' +2\mathbf{L}u' + \mathbf{M}u = 0$, using $u=\mathbf{R}v$ as
\begin{equation}
\begin{split}
v''+\mathbf{R}^{-1}(-\mathbf{L}^2-\mathbf{L}'+\mathbf{M})\mathbf{R}v&= v''+\mathbf{R}^{-1}\mathbf{Q}\mathbf{R}v\\
&=0, \label{veq}
\end{split}
\end{equation}
where the matrix, $\mathbf{Q}$ is labeled
\begin{eqnarray}
\mathbf{Q} = \left(\begin{array}{cc}A_Q & B_Q\\C_Q & D_Q\end{array}\right)
\end{eqnarray}
and the coefficients are given by
\begin{eqnarray}
\begin{split}
A_Q &= 3\eta_{\sigma\sigma}-6\epsilon+3\xi_1 s^2_\theta c_\theta + 10\epsilon\eta_{\sigma\sigma}-18\epsilon^2+ 11\epsilon\xi_1 s^2_\theta c_\theta\\ & \quad \quad -\eta_{\sigma\sigma}\xi_1 s^2_\theta c_\theta + \xi_1^2 s^4_\theta - \eta_{\sigma s}\xi_1 s_\theta(1+c_\theta^2),\label{Aeq}\\
B_Q &= 3\eta_{\sigma s} - 3\xi_1s_\theta^3 + 8\epsilon\eta_{\sigma s} - 9\epsilon\xi_1 s_\theta^3 + \eta_{\sigma\sigma}\xi_1 s_\theta^3 - \eta_{\sigma s}\xi_1 c_\theta^3+\xi_1^2s_\theta^3 c_\theta,\label{Beq}\\
C_Q &= 3\eta_{\sigma s} - 3\xi_1s_\theta^3 + 8\epsilon\eta_{\sigma s} - 9\epsilon\xi_1 s_\theta^3 + \eta_{\sigma\sigma}\xi_1 s_\theta^3 - \eta_{\sigma s}\xi_1 c_\theta^3+\xi_1^2s_\theta^3 c_\theta,\label{Ceq}\\
D_Q &= 3\eta_{ss} - 3\xi_1 c_\theta(1+s_\theta^2)+6\epsilon\eta_{ss} - 7\epsilon\xi_1 c_\theta(1+s_\theta^2)+ \eta_{\sigma\sigma}\xi_1 c_\theta(1+s_\theta^2)\\
& \quad\quad +\eta_{\sigma s}\xi_1 s_\theta c_\theta^2 + \xi_1^2(s_\theta^4-c_\theta^2) - \xi_2.\label{Deq}
\end{split}
\end{eqnarray}
Next, we diagonalise $\mathbf{Q}$ using
\begin{eqnarray}
\mathbf{R}_*^{-1}\mathbf{Q}\mathbf{R}_* = \left(\begin{array}{cc} \tilde{\lambda}_{1_*} & X \\ Y & \tilde{\lambda}_{2*}\end{array}\right), \label{rotation}
\end{eqnarray}
to find find an expression for $\tilde{\lambda}_*$ where $X = Y = 0$:
\begin{eqnarray}
\tilde{\lambda}_{1,2*} = \frac{A_{Q*}+D_{Q*}}{2} \pm \frac{A_{Q*}-D_{Q*}}{2}\sqrt{1+\frac{4B_{Q*}^2}{(A_{Q*}-D_{Q*})^2}} ,\label{lambdaeq}
\end{eqnarray}
or, more conveniently, to find a series of useful relations to be used later:
\begin{eqnarray}
\tilde{\lambda}_{1*}+\tilde{\lambda}_{2*} &=& A_{Q*} + D_{Q*}, \label{lambdaeq1} \\
(\tilde{\lambda}_{1*}-\tilde{\lambda}_{2*})\sin2\Theta_* &=& 2B_{Q*},\\
(\tilde{\lambda}_{1*}-\tilde{\lambda}_{2*})\cos2\Theta_* &=& A_{Q*}-D_{Q*},\\
\tilde{\lambda}_{1*}^2+\tilde{\lambda}_{2*}^2 &=& A_{Q*}^2+D_{Q*}^2+2B_{Q*}^2,\label{lambdaeq4}
\end{eqnarray}
Here,  in order to complete this diagonalisation, we have evaluated the rotation matrix at horizon crossing --- when $k = (aH)_*$ --- which shall be indicated by the subscript $_*$ from Eq.\ref{rotation} onwards. This approximation is justified by the fact that in this early evolution regime, up to and including the region around horizon crossing, the slow roll parameters are small and almost constant (as demonstrated later in the numerics) so our second order expansion will be more than enough to capture the dynamics.
Following \cite{Lalak07}, we perform another change of variable, $w = \mathbf{R_*^{-1}R_*}v$, to rewrite the above system of equations, Eq.(\ref{veq}), as two independent equations:
\begin{eqnarray}
w''_A + \left[k^2-\frac{1}{\tau^2}(2+3\lambda_{A*}) \right]w_A = 0 \label{fullweq}
\end{eqnarray}
whose solution is given by
\begin{eqnarray}
w_A = \frac{\sqrt{\pi}}{2}e^{i(\mu_A+1/2)\pi/2}\sqrt{-\tau}H_{\mu_A}^{(1)}(-k\tau)e_A(k),
\end{eqnarray}
where $H_{\mu_A}^{(1)}$ is a Hankel Function of the first kind, of order $\mu_A = \sqrt{\frac{9}{4}+3\lambda_{A*}}$ and the $e_A$ are two normalised Gaussian random variables.
In this new form, it should be noted that we have moved from $\tilde{\lambda_*}$ to $\lambda_*$ through the following relation --- this now takes into account the $\frac{a''}{a}$ term omitted earlier, in Eq.~(\ref{adoubledash})
\begin{eqnarray}
\lambda_{A*} = \epsilon_* + \frac{20\epsilon^2_*}{3}-\frac{8\epsilon_*\eta_{\sigma\sigma*}}{3} - \frac{8\epsilon_*\xi_{1*} s^2_{\theta*} c_{\theta*}}{3} -\frac{\tilde{\lambda}_{A*}}{3}
\end{eqnarray}
and we have expanded $\mu_A \approx \frac{3}{2}+\lambda_{A*}-\frac{\lambda_{A*}^2}{3} $ to second order in the slow roll parameters.
Using the regular definition of the power spectra 
\begin{equation}
\left<Q_{A}({\bf k})Q_B({\bf k'})\right> = 8\pi^3\delta^{(3)}({\bf k}+{\bf k'})\frac{2\pi^2}{k^3}\mathcal{P}_{AB}(|{\bf k}|),
\end{equation}
and the following relations, coming from the earlier rotations
\begin{eqnarray}
a^2\left<Q^\dagger_\sigma Q_\sigma\right> &=& \cos^2\Theta_*\left<w^\dagger_1 w_1\right> + \sin^2\Theta_*\left<w^\dagger_2 w_2\right>,\label{weq1}\\
a^2\left<\delta s^\dagger \delta s\right> &=& \sin^2\Theta_*\left<w^\dagger_1 w_1\right> + \cos^2\Theta_*\left<w^\dagger_2 w_2\right>,\\
a^2\left<\delta s^\dagger Q_\sigma\right> &=& \frac{\sin2\Theta_*}{2}\left(\left<w^\dagger_1 w_1\right>-\left<w^\dagger_2 w_2\right> \right),\label{weq3}
\end{eqnarray}
we see that we need to find the correlation functions associated with $w_{A}$, where we relabel $-k\tau=x$:
\begin{equation}
\begin{split}
\left<w_A^\dagger w_A\right> &= \frac{-\tau \pi}{4} \mid H_{\mu_A*}^{(1)}(-k\tau)\mid^2 \\
&= \frac{-\tau \pi}{4}\left(1+2\lambda_{A*} f(x)+\lambda_{A*}^2 g(x) \right)\mid H_{\frac{3}{2}*}^{(1)}(x)\mid^2\\
&= \frac{-\tau}{2x^3}\left(1+2\lambda_{A*} f(x)+\lambda_{A*}^2 g(x) \right)(1+x^2).\label{eqww}
\end{split}
\end{equation}
This comes from the Taylor expansion of the Hankel function around the point $\mu = \frac{3}{2}$ along with evaluating the Hankel function itself at that point. For brevity, in the above we have used  
\begin{align}
f(x) &= {\rm Re}\left[\left.\frac{1}{H^{(1)}_{3/2}(x)} \frac{dH_{\mu}^{(1)}(x)}{d\mu}\right|_{\mu=3/2} \right]  \quad \text{and}\\
g(x) &= {\rm Re}\left[ \left.\frac{1}{H^{(1)}_{3/2}(x)} \frac{d^2H_{\mu}^{(1)}(x)}{d\mu^2}\right|_{\mu=3/2} + \left(\left.\frac{1}{H^{(1)}_{3/2}(x)} \frac{dH_{\mu}^{(1)}(x)}{d\mu}\right|_{\mu=3/2} \right)^2-\frac{2}{3}\left.\frac{1}{H^{(1)}_{3/2}(x)} \frac{dH_{\mu}^{(1)}(x)}{d\mu}\right|_{\mu=3/2}\right]
\end{align}
These functions can be evaluated at $x \rightarrow 0$ as:
\begin{align}
f(x) &= 2-\gamma-\ln{2} -\ln{x}\\
6g(x)&= 16+3\pi^2-44\gamma+12\gamma^2+24\gamma\ln{2}-44\ln{2}+12\ln^2{2} \notag\\
&\quad\quad + 12\ln^2{x} -44\ln{x}+24\gamma\ln{x}+24\ln{x}\ln{2} \label{6gxeq}
\end{align}
By inserting these results into Eq's~(\ref{weq1})-(\ref{weq3}) and expanding the scale factor, $a$, to second order we can calculate the final power spectra in the $\sigma, \delta s$ basis. It is then useful to convert these results back to the more common comoving curvature perturbation and associated renormalised entropy perturbation using the following relations, respectively
\begin{eqnarray}
\mathcal{R} &\equiv& \frac{H}{\dot{\sigma}}Q_\sigma,\\
\mathcal{S} &\equiv& \frac{H}{\dot{\sigma}}\delta s.
\end{eqnarray}
The final results have been written both in terms of the matrix coefficients, $A_{Q*},B_{Q*},C_{Q*}$ and $D_{Q*}$, used earlier as a natural result of expanding the trigonometric identities involving $\lambda_*$ (Eq's (\ref{lambdaeq1})-(\ref{lambdaeq4})) and in terms of the slow roll parameters (also evaluated at horizon exit) directly. Whilst this mix is unfortunate, the equations can be kept reasonably concise if the coefficients are not directly expanded in terms of their slow roll parameters here---the results of which can be found in Eq.~(\ref{Deq}).
\begin{equation}
\begin{split}
\mathcal{P}_{\mathcal{R}} &= \frac{H^2_*}{8\pi^2\epsilon_*} (1-2\epsilon_*-11\epsilon_*^2+4\epsilon_*\eta_{\sigma\sigma*}+4\epsilon_*\xi_{1*} s^2_{\theta*} c_{\theta*})(1+k^2\tau^2) \times \\
& \quad\quad\quad\left[1+ \frac{2}{3}(3\epsilon_*+20\epsilon_*^2-8\epsilon_*\eta_{\sigma\sigma*}-8\epsilon_*\xi_{1*} s^2_{\theta*} c_{\theta}* -A_{Q*})f(x) \right.\\
& \quad\quad\quad\quad \left.\ + \left(\epsilon_*^2+\frac{A_{Q*}^2+B_{Q*}^2}{9}-\frac{2\epsilon_* A_{Q*}}{3} \right) g(x) \right],
\end{split}
\end{equation}
\begin{equation}
\begin{split}
\mathcal{P}_{\mathcal{S}} &= \frac{H^2_*}{8\pi^2\epsilon_*} (1-2\epsilon_*-11\epsilon_*^2+4\epsilon_*\eta_{\sigma\sigma*}+4\epsilon_*\xi_{1*} s^2_{\theta*} c_{\theta*})(1+k^2\tau^2) \times \\
& \quad\quad\quad\left[1+ \frac{2}{3}(3\epsilon_*+20\epsilon_*^2-8\epsilon_*\eta_{\sigma\sigma*}-8\epsilon_*\xi_{1*} s^2_{\theta*} c_{\theta}* -D_{Q*})f(x) \right.\\
& \quad\quad\quad\quad \left.\ + \left(\epsilon_*^2+\frac{D_{Q*}^2+B_{Q*}^2}{9}-\frac{2\epsilon_* D_{Q*}}{3} \right) g(x) \right],
\end{split}
\end{equation}
\begin{equation}
\begin{split}
\mathcal{C}_{\mathcal{RS}} &= \frac{H^2_*}{8\pi^2\epsilon_*} (1-2\epsilon_*-11\epsilon_*^2+4\epsilon_*\eta_{\sigma\sigma*}+4\epsilon_*\xi_{1*} s^2_{\theta*} c_{\theta*})(1+k^2\tau^2) \times \\
& \quad\quad\quad\left[\frac{B_{Q*}}{9}(A_{Q*}+D_{Q*}-6\epsilon_*)g(x) - \frac{2}{3}B_{Q*} f(x)\right].
\end{split}
\end{equation}

\subsection{Evolution on super-Hubble scales}
If the two equations held within Eq.~(\ref{exacteq}) are combined, to first order in slow roll we obtain
\begin{eqnarray}
\frac{\dot{Q_\sigma}}{H} &=& -(\eta_{\sigma\sigma}-2\epsilon + \xi_1c_\theta s_\theta^2)Q_\sigma - 2(\eta_{\sigma s}-\xi_1s_\theta^3)\delta s, \label{qsigmadot}\\
\frac{\dot{\delta s}}{H} &=& -\left(\eta_{ss}-\xi_1c_\theta(1+s_\theta^2)\right)\delta s \label{deltasdot},
\end{eqnarray}
since on super-Hubble scales we can neglect both the $\frac{k^2}{a^2}$ term, as the wavenumber is now small, and the double time derivatives which change slowly in comparison to the scale factor. Following \cite{Davis12}, by differentiating these with respect to time it is possible to calculate second order expressions in terms of the slow roll parameters, using the expansions found in appendix A, for $\ddot{Q_\sigma}$ and $\ddot{\delta s}$.  Finally, these can then be inserted back into Eq.~(\ref{exacteq}) to find second order expressions for the full equations of motion:\\

\begin{eqnarray}
\frac{\dot{Q}_\sigma}{H} &=& AQ_\sigma + B\delta s \label{finalQ1},\\
\frac{\dot{\delta s}}{H} &=& D\delta s\label{finalds1},
\end{eqnarray}
where
\begin{equation}
\begin{split}
A &= \left(2\epsilon - \eta_{\sigma\sigma}-\xi_1 s_\theta^2 c_\theta-\frac{\eta_{\sigma s}^2}{3} -\frac{4\epsilon^2}{3}-\frac{\eta_{\sigma\sigma}^2}{3} +\frac{5\epsilon\eta_{\sigma\sigma}}{3} -\frac{2\xi_1^2 s^2_\theta c^2_\theta}{3}+ \frac{\xi_2 s^2_\theta c^2_\theta}{3}\right.\\
& \quad \left.  -\frac{4\eta_{\sigma\sigma}\xi_1 s^2_\theta c_\theta}{3}- \frac{4\eta_{\sigma s}\xi_1 s_\theta c_\theta^2}{3}+\frac{4\epsilon\xi_1 s^2_\theta c_\theta}{3} -\frac{\alpha_{\sigma\sigma\sigma}}{3}\right),
\end{split}
\end{equation}
\begin{equation}
\begin{split}
B &= \left(-2\eta_{\sigma s} +2\xi_1 s^3_\theta + 2\epsilon\eta_{\sigma s}-\frac{2\eta_{\sigma\sigma}\eta_{\sigma s}}{3}-\frac{2\eta_{ss}\eta_{\sigma s}}{3}+\frac{4\eta_{\sigma\sigma}\xi_1 s^3_\theta}{3}- \frac{4\epsilon\xi_1 s^3_\theta}{3}\right.\\
& \quad \left. - \frac{4\eta_{ss}\xi_1s_\theta c^2_\theta}{3}+ \frac{4\xi_1^2s^3_\theta c_\theta}{3}- \frac{2\xi_2 c_\theta s_\theta^3}{3}- \frac{2\alpha_{\sigma\sigma s}}{3}\right),
\end{split}
\end{equation}
\begin{equation}
\begin{split}
D &= \left(-\eta_{ss}+\xi_1c_\theta(1+s_\theta^2)-\frac{\eta_{\sigma s}^2}{3}-\frac{\eta_{ss}^2}{3}+\frac{\epsilon\eta_{ss}}{3}-\frac{\alpha_{\sigma ss}}{3} + \frac{\xi_2 s_\theta^4}{3}+\frac{4\eta_{\sigma s}\xi_1 s_\theta^3}{3} \right. \\
&\quad \left. - \frac{4\xi_1^2 s_\theta^4}{3} + \frac{4\eta_{ss}\xi_1 c_\theta s_\theta^2}{3}\right),
\end{split}
\end{equation}\label{Geq}
which can then be integrated and used to find the final power spectra as \cite{Lalak07}
\begin{eqnarray}
\mathcal{P_R}(N) &=& \mathcal{P_R}_* \left[ 1 + \left(\displaystyle \int^{N}_{N_*} B(N'')e^{\int^{N}_{N_*}\gamma(N')dN'} dN'' \right)^2 \right. \nonumber  \\
 &-& \left. 2\eta_{\sigma s} f(-k \tau_*) \displaystyle \int^{N}_{N_*} B(N'') e^{\int^{N}_{N_*}\gamma(N')dN'} dN''  
 \right] ,\\
\mathcal{P_S}(N) &=& \mathcal{P_S}_*e^{2\int^{N}_{N_*}\gamma(N')dN'},
\end{eqnarray}
where the `$_*$' denotes the value of the power spectra at horizon crossing and $\gamma = D - A$.

\section{Numerical method and models}
In the remainder of the paper, we are going to compare our expressions for power spectra against exact numerical calculations. In this section we discuss the numerical methods employed and give examples of the trajectories studied later. The perturbation equations (\ref{exacteq}) are integrated twice, independently, where in the first instance we set the initial value of $\delta s$ to zero and the initial value of $Q_\sigma$ to that of the Bunch-Davies vacuum and vice versa for the second run. The two runs are then combined to find the power spectra via 
\begin{eqnarray}
\mathcal{P_R} &=& \frac{k^3}{2\pi^2}\left(\lvert\mathcal{R}_1\rvert^2 + \lvert\mathcal{R}_2\rvert^2 \right) \quad \text{and} \\
\mathcal{P_S} &=& \frac{k^3}{2\pi^2}\left(\lvert\mathcal{S}_1\rvert^2 + \lvert\mathcal{S}_2\rvert^2 \right)
\end{eqnarray}
where
\begin{eqnarray}
\mathcal{R}^2_1 &=& \left((Q_{\sigma 1}^{Re})^2+ (Q_{\sigma 1}^{Im})^2\right)/\sigma^{' 2},\\
\mathcal{R}^2_2 &=& \left((Q_{\sigma 2}^{Re})^2+ (Q_{\sigma 2}^{Im})^2\right)/\sigma^{' 2},\\
\mathcal{S}^2_1 &=& \left((\delta s^{Re}_1)^2+ (\delta s^{Im}_1)^2\right)/\sigma^{' 2},\\
\mathcal{S}^2_2 &=& \left((\delta s^{Re}_2)^2+ (\delta s^{Im}_2)^2\right)/\sigma^{' 2}.\\
\end{eqnarray}
In these expressions the prime denotes differentiation with respect to the e-fold number, $N$.

In this paper we choose an appropriate wave--number $k$ such that the mode exits 8 e-folds after the start of the integration, which makes the comparison with earlier work easier. The results of both the analytical work and numerics are then normalised by dividing through by the single field result, $\mathcal{P}_* = \frac{H^2_*}{8\pi^2\epsilon}$. It should also be noted that the point of transition between the  horizon crossing regime and the subsequent evolution is somewhat arbitrary --- as there is no point at which one completely breaks down whilst the other remains useful. Rather, the region of accurate overlap is suitably large for the precise point to be chosen quite loosely, with checks made afterwards.

\subsection{Models and background trajectories}

\begin{figure}[H]
  \includegraphics[width=\linewidth]{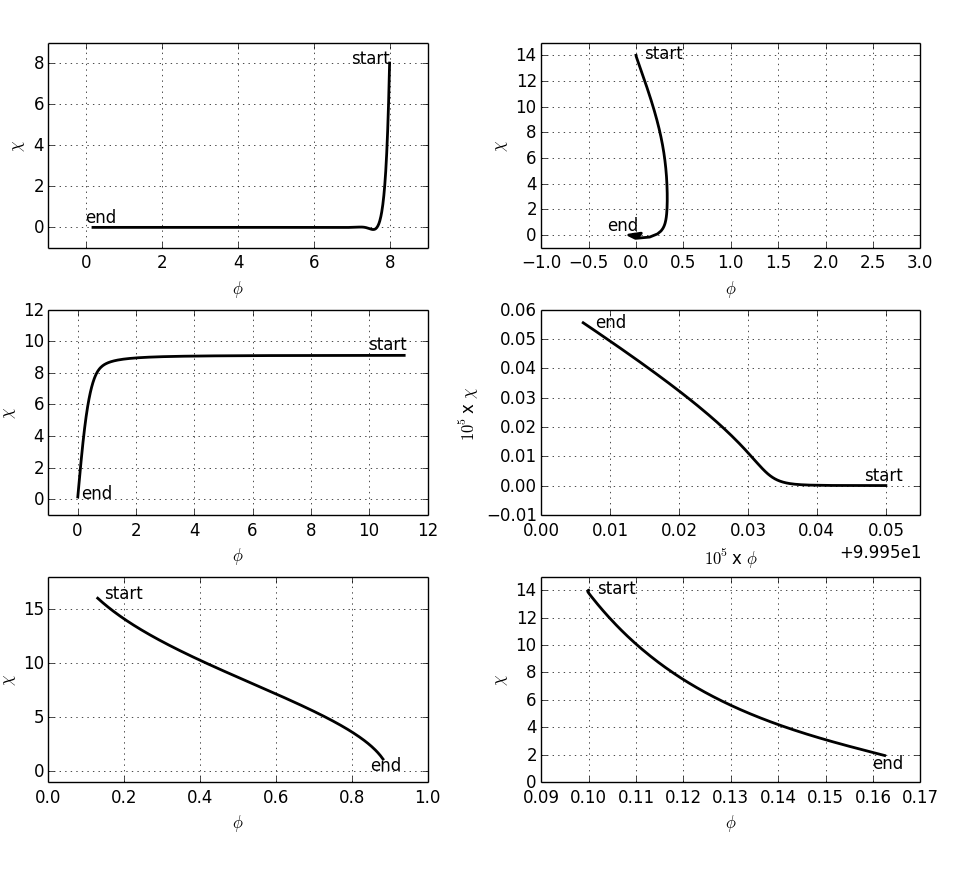}
  \caption{Inflationary trajectories. The top left panel shows the double quadratic potential with $b(\phi) = 0$ and the top right shows it with $b(\phi) = \phi$. The middle row shows the quartic potential on the left and the hybrid evolution on the right. Finally the bottom row shows the product potential trajectory on the left and, on the right, the second order non-canonical term case of double quadratic inflation --- so $b(\phi) = \phi^2$.}
  \label{bgtraj}
\end{figure}

We look at a number of variations (by varying the parameter, $b(\phi)$) based upon the following four models: a double quadratic potential, a quartic potential, a product potential and hybrid inflation. The precise form of each is as follows, and some example background trajectories of the fields can be seen in figure \ref{bgtraj}:

\begin{itemize}
  \item Double Inflation
\end{itemize}
\begin{equation}
V(\phi,\chi) = \frac{1}{2}m_\phi^2\phi^2+\frac{1}{2}m_\chi^2\chi^2,
\end{equation}
\begin{itemize}
  \item Quartic Inflation
  \end{itemize}
\begin{equation}
V(\phi,\chi) = \frac{\lambda_\phi \phi^4}{4} + \frac{\lambda_\chi \chi^4}{4},
\end{equation}
where we follow \cite{Ash10} in using $\lambda_\phi/\lambda_\chi = 410$, $\phi_0 = 11.2$ and $\chi_0 = 9.1$.
\begin{itemize}
  \item Hybrid Inflation
  \end{itemize}
  \begin{equation}
V(\phi,\chi)=\Lambda^4\left[\left(1-\frac{\chi^2}{v^2} \right)^2+\frac{\phi^2}{\mu^2}+\frac{2\phi^2\chi^2}{\phi^2_C v^2} \right],
\end{equation}
where we follow \cite{Kodama11} by setting $\nu = 0.10, \phi_c = 0.01, \mu = 1000$ and we can choose the normalisation constant, $\Lambda$ arbitrarily. This equates to a variant of hybrid inflation motivated by \cite{Clesse11,Abo11} which avoids the blue spectral index commonly associated with earlier hybrid models. 
\begin{itemize}
  \item Product Inflation
\end{itemize}
\begin{equation}
V(\phi, \chi) = \frac{m_\chi^2 \chi^2}{2} e^{-\lambda \phi^2},
\end{equation}\label{prodpot}
motivated by example A in \cite{Byrnes08}, in which case we take $\phi_0 = 0.13$, $\chi_0 = 16$ and $\lambda = 0.015$.

\section{Numerical results}

Here we discuss the results of the numerical simulations alongside those achieved analytically, including comparisons to earlier work and the improvements found here. Following convention, we associate the zeroth e-fold with the time that the mode of interest exits the horizon so the sub-horizon evolution is denoted by negative $N$. 

\subsection{Double Inflation}

\subsubsection{$b(\phi) = 0$}

\begin{figure}
  \includegraphics[width=\linewidth]{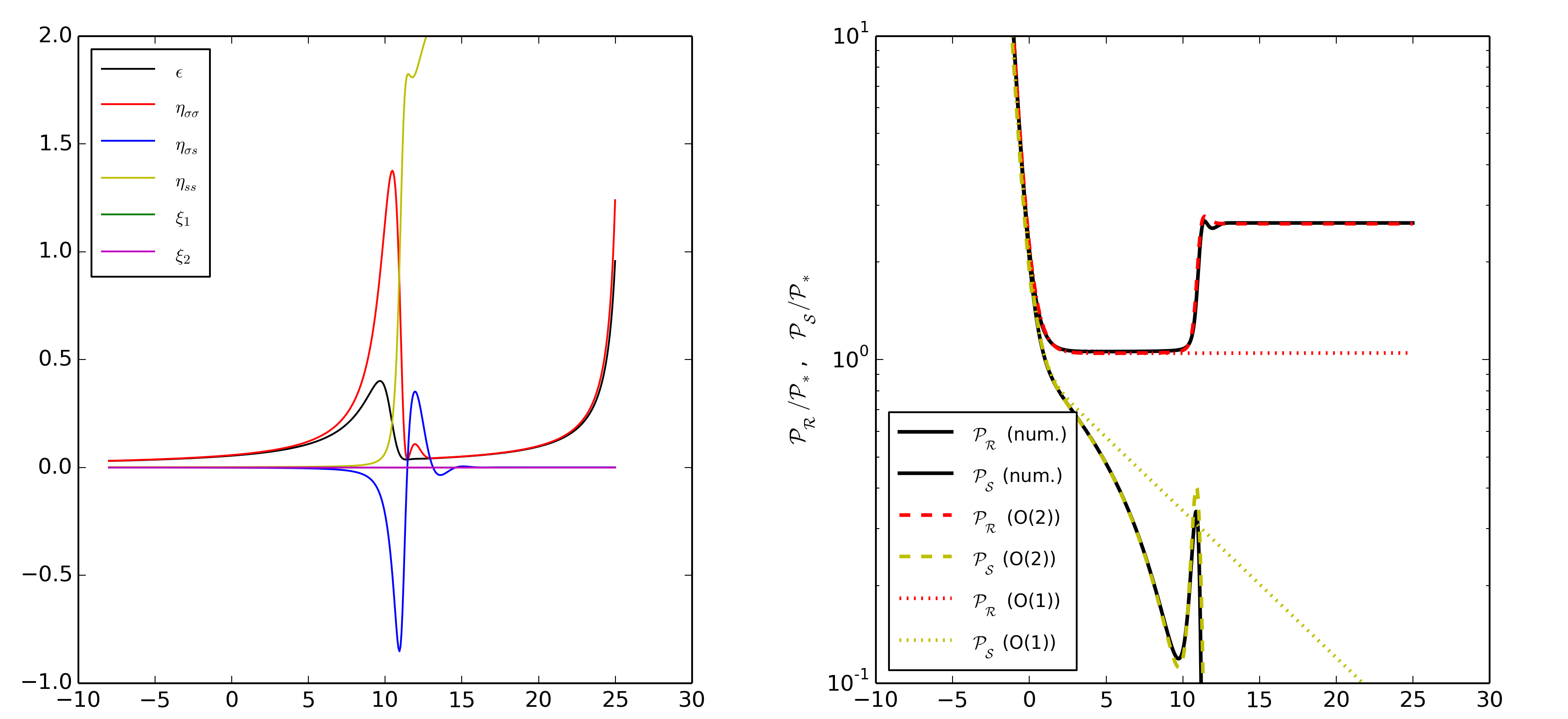}
  \caption{Canonical double inflation with a mass ratio of 7 ( $m_\chi/m_\phi = 7$). The left graph shows the evolution of the slow roll parameters for canonical double inflation. The right graph shows the resulting power spectra with the solid black lines representing the numerical results, the dotted lines representing the first order results and the dashed lines representing the second order results.}
  \label{candoub7}
\includegraphics[width=\linewidth]{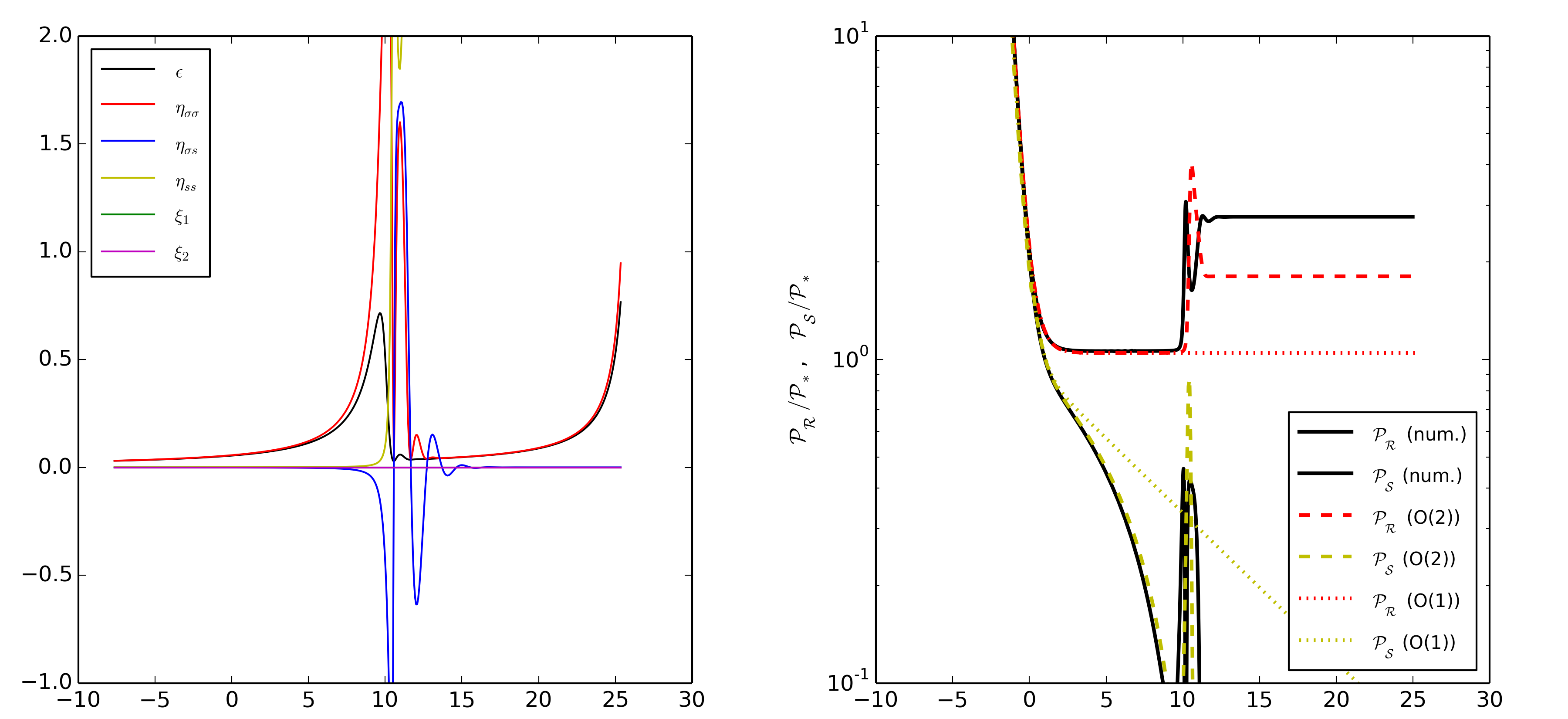}
  \caption{Canonical double inflation with $m_\chi/m_\phi = 9$. The left graph shows the evolution of the slow roll parameters. The right graph shows the resulting power spectra with the solid black lines representing the numerical results, the dotted lines representing the first order results and the dashed lines representing the second order results.}
  \label{candoub10}
\end{figure}

The first example uses $b(\phi) = 0$ which leaves us with the standard, non-coupled case. As can be seen in the top left graph in figure \ref{bgtraj}, inflation begins with the $\chi$ field dominant before a sharp turn in field space leaves the field rolling down in the $\phi$ direction. During each of these phases --- where one field dominates over the other --- we see few interesting effects as isocurvature perturbations are minimal. However, over the transition period we find significant isocurvature perturbations feeding the curvature perturbations which accounts for the large increase in $\mathcal{P_R}$. We find, in agreement with \cite{Davis12}, that our second order approximation captures this effect well for field mass ratios up to $\frac{m_\chi}{m_\phi} \sim 9$. With the example shown in figure \ref{candoub7} using $\frac{m_\chi}{m_\phi} \sim 7$. If we go beyond this ratio, however, we start to see higher order effects play a role as the subsequent `wobbles' as the field settles down after the initial turn in field space allow numerous additional opportunities for the isocurvature perturbations to feed the curvature. This is evident in figure \ref{candoub10} where the final power spectra is underestimated. Note that both the curvature and isocurvature perturbations show discrepancies to the full numerical result. For this limiting case our analytical expressions underestimate the sourcing of the curvature perturbation from the isocurvature perturbation.

In both of these cases we see, on the left hand side of figures \ref{candoub7} and \ref{candoub10}, that the slow roll parameters are no longer small during this transition period --- with both $\eta_{ss}$ and $\eta_{\sigma \sigma}$ becoming, or exceeding, unity. This is especially true in the higher mass ratio case which explains why there is such a significant difference between the analytical and numerical results. In both cases this behavior soon settles down again during the second phase allowing for further inflation, although it should be noted that $\eta_{ss}$ does remain large throughout.

\subsubsection{$b(\phi) = \phi$}

\begin{figure}
\includegraphics[width=\linewidth]{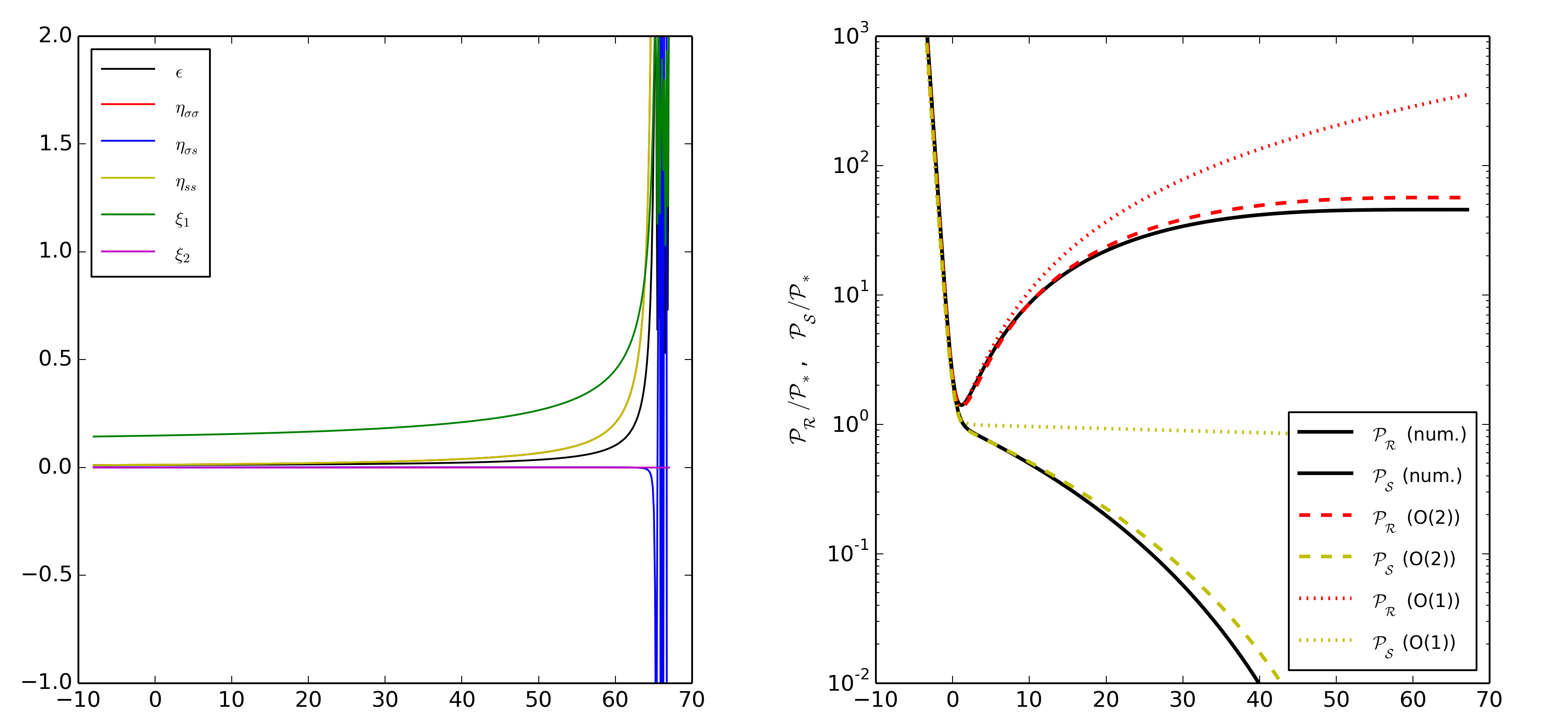}
  \caption{Non-canonical double inflation with $m_\chi/m_\phi = 7$. The left graph shows the evolution of the slow roll parameters - with $\xi_1$ dominating. The right graph shows the resulting power spectra with the solid black lines representing the numerical results, the dotted lines representing the first order results and the dashed lines representing the second order results.}
  \label{noncandoub}
\includegraphics[width=\linewidth]{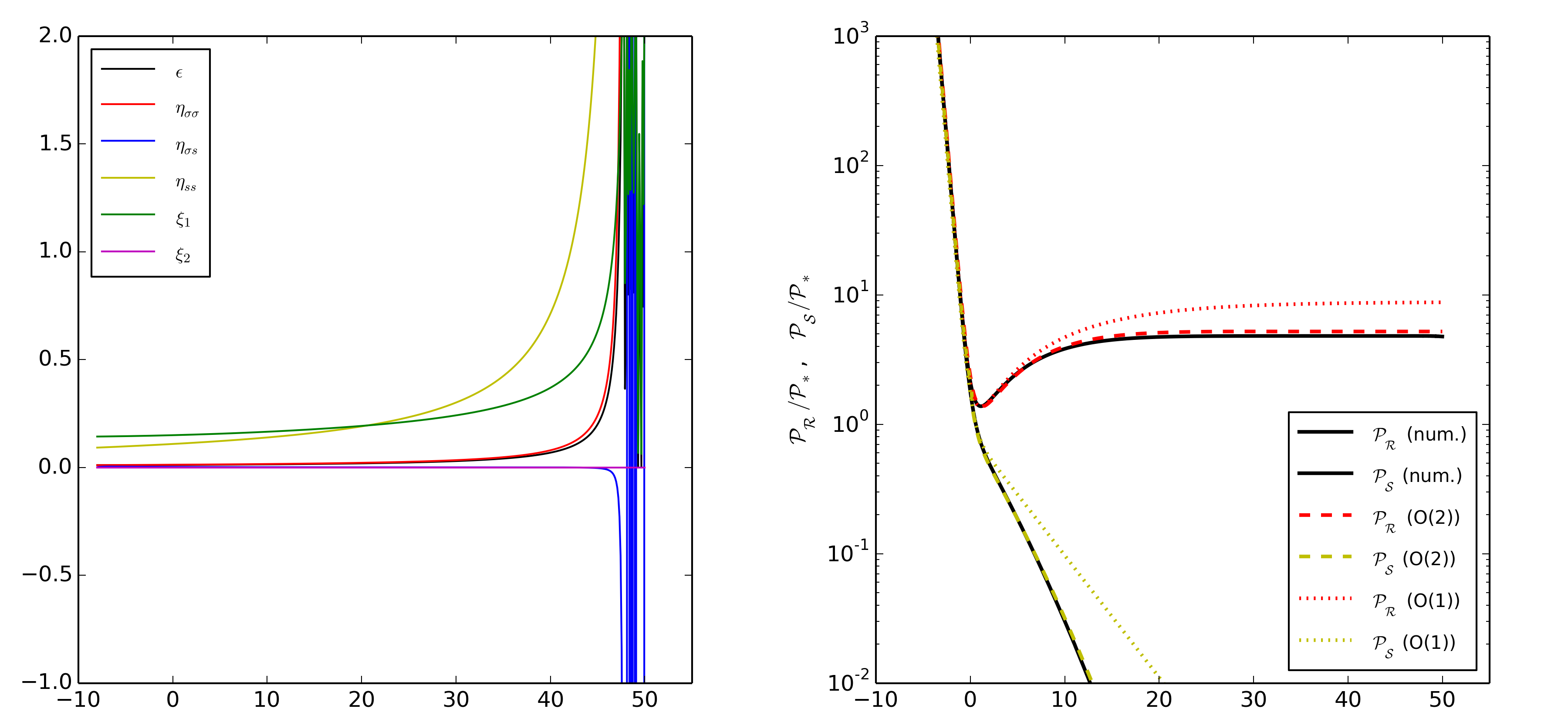}
  \caption{Non-canonical double inflation with $m_\chi/m_\phi = 1/3$. The left graph shows the evolution of the slow roll parameters - now with both $\xi_1$ and $\eta_{ss}$ dominating. The right graph shows the resulting power spectra with the solid black lines representing the numerical results, the dotted lines representing the first order results and the dashed lines representing the second order results.}
  \label{noncandoub2}
\end{figure}

Before looking at a quadratic form for $b(\phi)$ for which  the new parameter, $\xi_2$, is non-zero (section 4.5), we first look at the linear non-canonical case.  Due to the form of our coupling term, in order to keep $\xi_1$ small we must choose a trajectory in which $\phi$ never gets too large, as shown in the top right of figure \ref{bgtraj}. We can see straight away that the effect of $b(\phi)$ is to pull the inflaton away from the $\phi$  minimum before coming back to settle at the true minimum of the potential at the end of inflation. In this instance it is evident that from the outset the non-canonical parameter, $\xi_1$, plays a dominant role by having a value of roughly $0.14$ even during horizon exit (in comparison to $\sim 0$ for all other slow roll parameters), as shown in figure \ref{noncandoub}. It is the size of this parameter that justifies our initial choice of $\phi_0 = 0$ as larger values would soon result in instances where $\xi_1 > 1$ for a prolonged period of time, thereby rendering the slow roll expansion insufficient. We can, however, vary the mass ratio, $\frac{m_\chi}{m_\phi}$ and see that, in general, we find much better agreement with the numerics than was found up to first order even with `large' $\xi_1$ terms throughout the evolution. There are exceptions in terms of $\xi_1$ dominating throughout, and that comes in the cases where the mass ratio, $m_\chi/m_\phi$, is a fraction, such as $\frac{m_\chi}{m_\phi} = \frac{1}{3}$ (figure \ref{noncandoub2}). In this case the motion is suppressed in the $\phi$ direction --- leading to an equal contribution from $\eta_{ss}$ alongside $\xi_1$. Although, due to the suppression the overall effect of this is small --- as the potential begins to resemble a single field case again in which the inflaton has little room to turn thereby minimising isocurvature-curvature mixing, which in turn minimises the super horizon contributions to the  power spectra.

\subsection{Quartic Potential}

\begin{figure}
\begin{center}
\includegraphics[width=\linewidth]{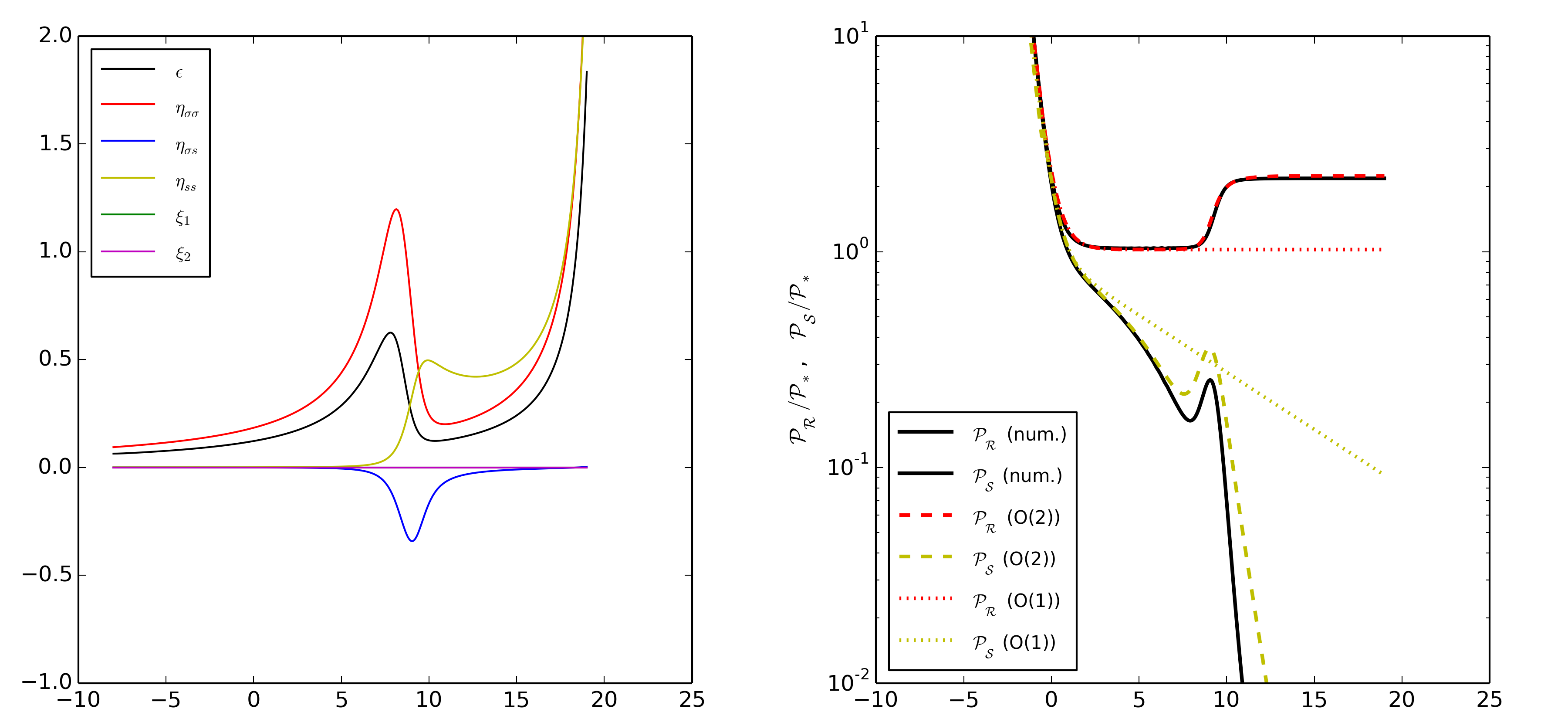}
  \caption{Quartic potential. The left graph shows the evolution of the slow roll parameters. The right graph shows the resulting power spectra with the solid black lines representing the numerical results, the dotted lines representing the first order results and the dashed lines representing the second order results.}
  \label{quarticps}
  \end{center}
\end{figure}

With such a similar potential to the double quadratic already discussed, and a similar trajectory in field space (figure \ref{bgtraj}) we do not expect much in the way of additional interest here. It is, however, worthy of inclusion to note the differences seen between our approximation and that of \cite{Davis12}. Whilst the previous publication noted significant discrepancies between not only the second order results, but also third order, in comparison to the amplitude of the numerical $\mathcal{P_R}$ --- we find a much closer fit with our full second order expansion in the canonical case. This contrasts with the isocurvature amplitude in which our approximation appears to overestimate the true value before quickly decaying before the end of inflation. These results are demonstrated in figure \ref{quarticps}.  

\subsection{Hybrid Inflation}

\begin{figure}
\includegraphics[width=\linewidth]{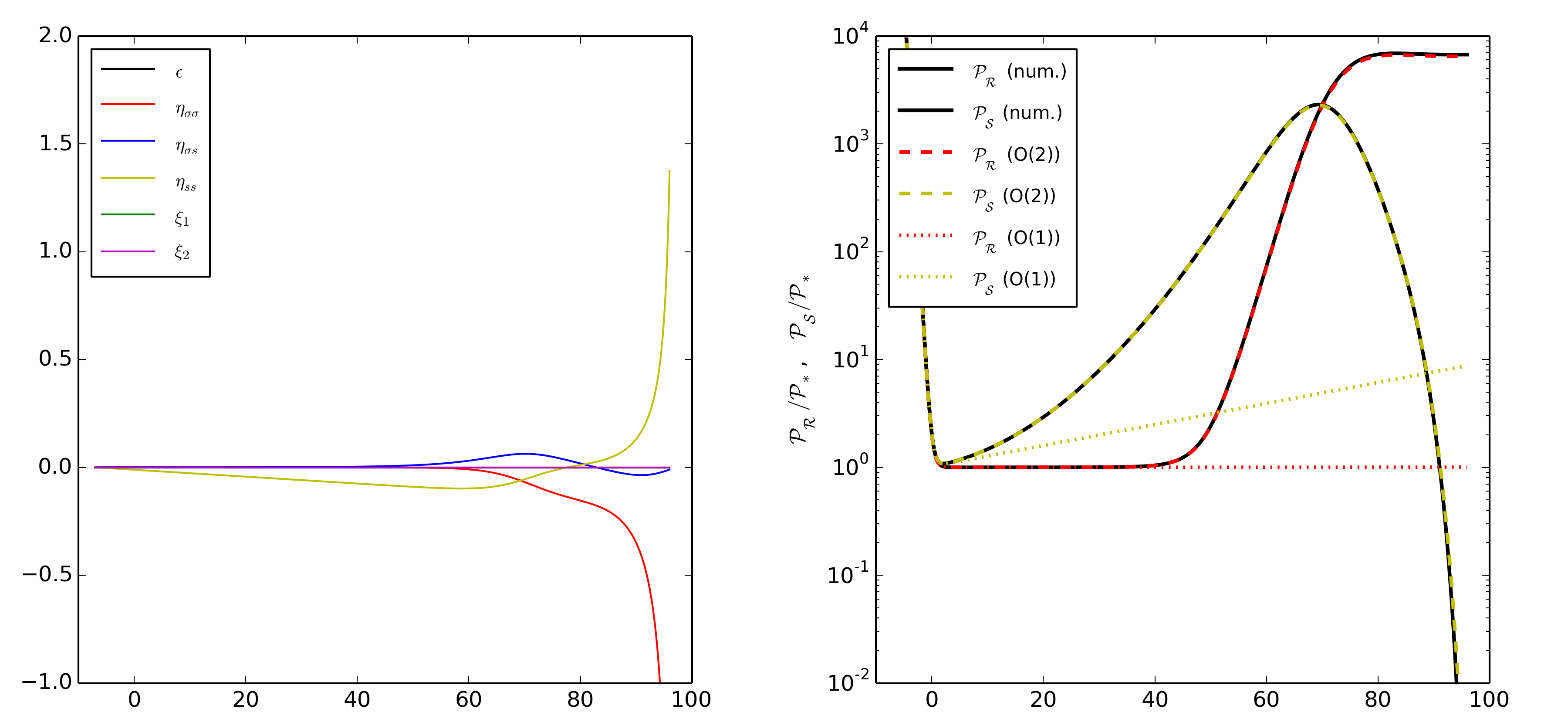}
  \caption{Canonical hybrid inflation. The left graph shows the evolution of the slow roll parameters --- which all remain relatively small. The right graph shows the resulting power spectra with the solid black lines representing the numerical results, the dotted lines representing the first order results and the dashed lines representing the second order results.}
  \label{canhybrid}
\includegraphics[width=\linewidth]{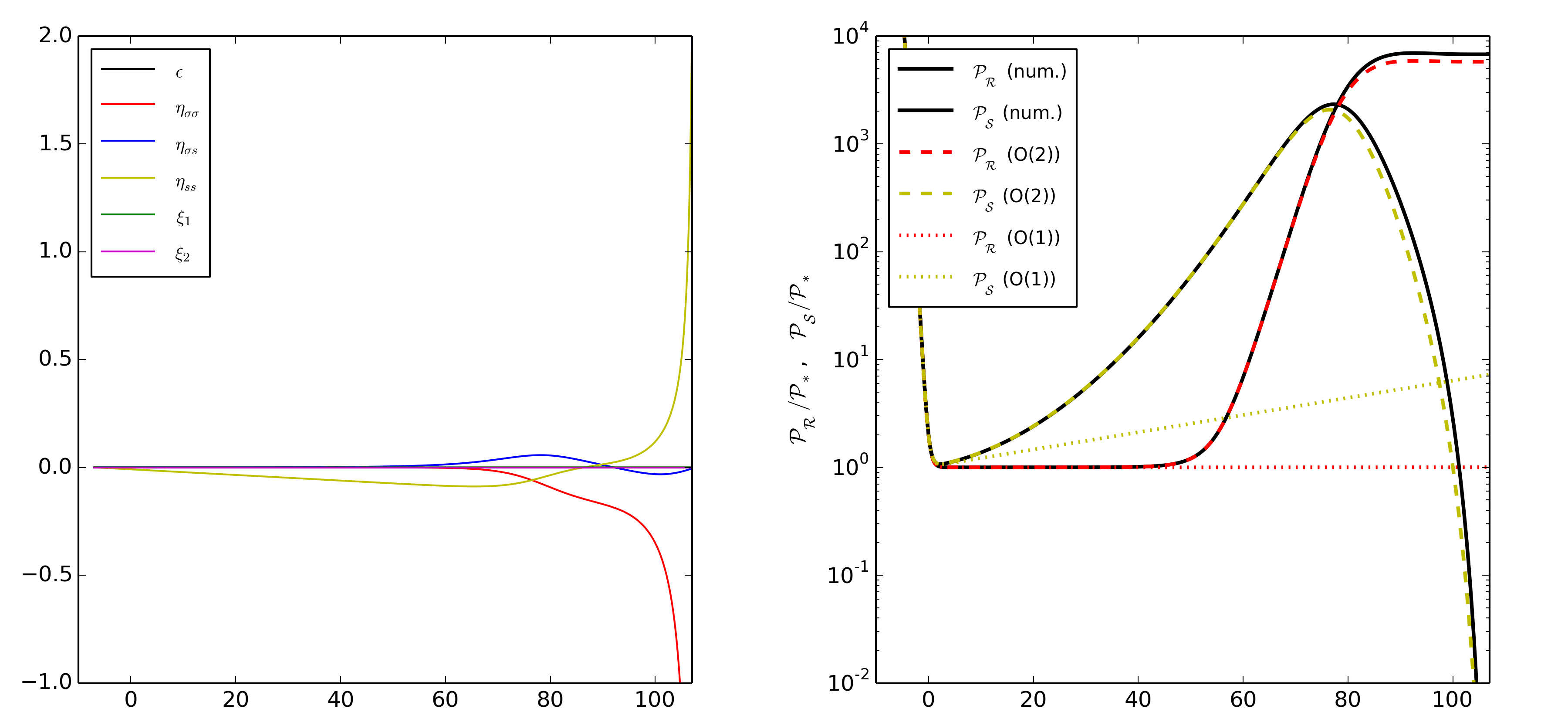}
  \caption{Non-canonical hybrid inflation with $\beta = 10$. The left graph shows the evolution of the slow roll parameters --- which, again, all remain relatively small. The right graph shows the resulting power spectra with the solid black lines representing the numerical results, the dotted lines representing the first order results and the dashed lines representing the second order results.}
  \label{noncanhybrid}
\end{figure}

Using this substantially different potential we test the accuracy of our approximation both in the canonical and non-canonical case. In the canonical case we find, as expected, that our approximation almost overlaps the numerical results. In the non-canonical case --- in which we use $b(\phi) = 10\phi$ --- we find a slight deviation, but this is still considerably better than the first order approximation which doesn't track the super horizon evolution at all. Although appearing similar graphically, the final amplitude in the non-canonical case is increased by a couple of percent in comparison to the canonical case, along with the evolution being slowed by up to 15\%. We choose a rather large value of $\beta = 10$ here due to the smallness of $\phi$ in such a model since using $\beta$ values similar to those in the double quadratic model would result in negligible change to the results. 

\subsection{Product potential}
\begin{figure}
\includegraphics[width=\linewidth]{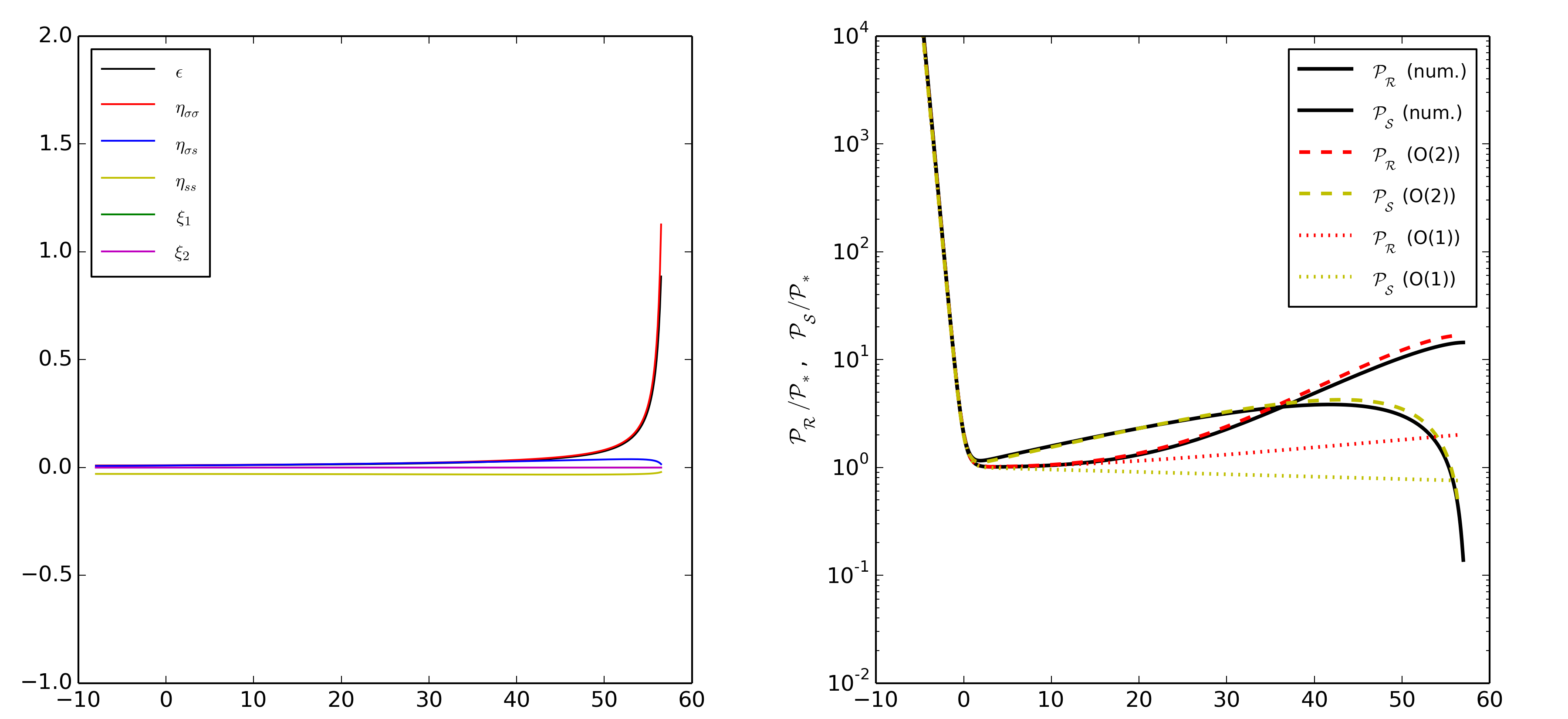}
  \caption{Product potential. The left graph shows the evolution of the slow roll parameters. The right graph shows the resulting power spectra with the solid black lines representing the numerical results, the dotted lines representing the first order results and the dashed lines representing the second order results.}
  \label{canprodpot}
\includegraphics[width=\linewidth]{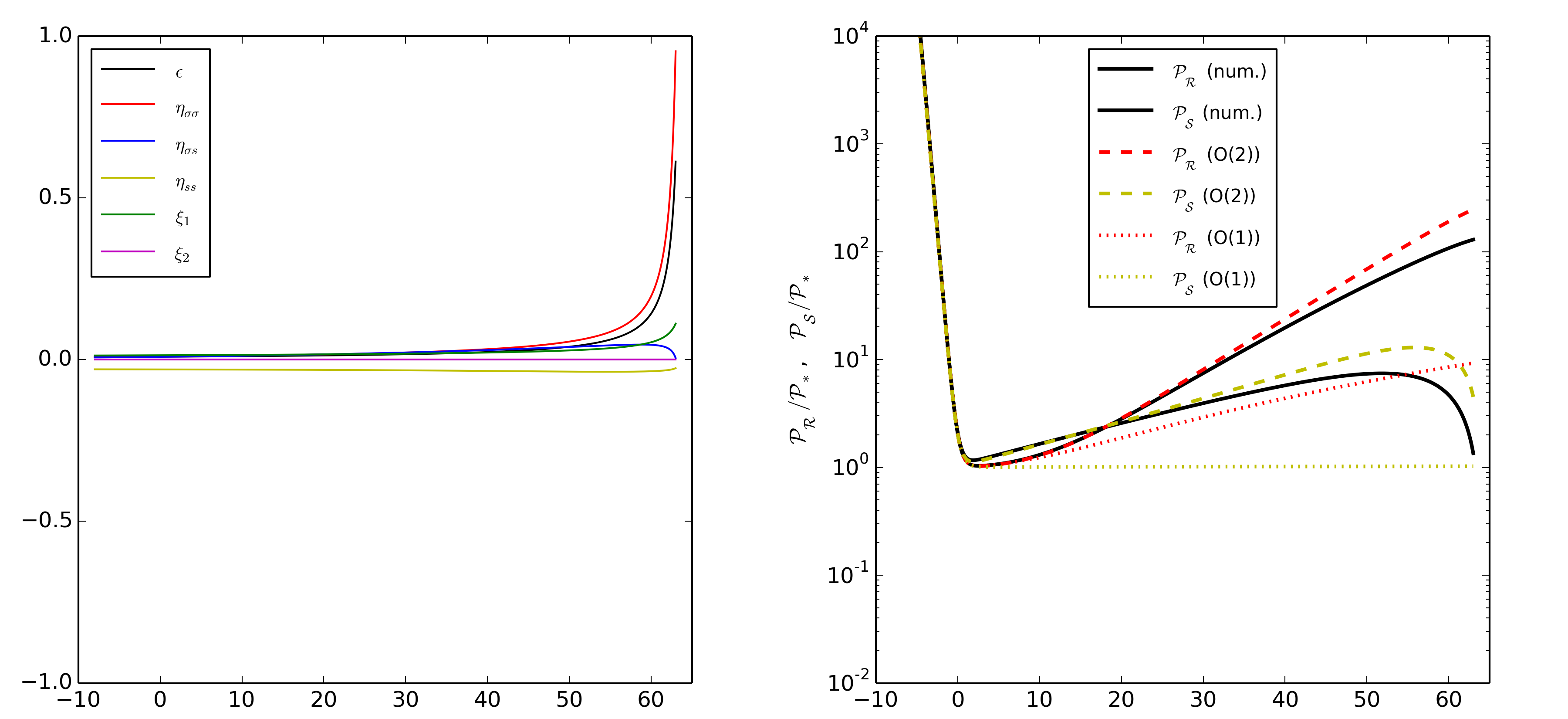}
  \caption{Non-canonical product potential, with $b(\phi) = 0.1\phi$. The left graph shows the evolution of the slow roll parameters. The right graph shows the resulting power spectra with the solid black lines representing the numerical results, the dotted lines representing the first order results and the dashed lines representing the second order results.}
  \label{noncanprodpot}
\end{figure}

Next we come to the product potential given in equation (\ref{prodpot}) with a trajectory shown in the bottom left of figure \ref{bgtraj}. In this case we demonstrate some limitations of our expansion that were also present in some previous results, but are much more apparent here. Figure \ref{canprodpot} again shows good agreement in the canonical case. Whilst not being quite as accurate as some previous models (in which the analytical results overlap with the numerics) the general form of the super-horizon evolution is present and reasonably accurate --- capturing the growth, sourcing and subsequent decay of the isocurvature perturbations --- whilst the usefulness of the first order version remains fairly limited. If we now include a non-canonical contribution, figure \ref{noncanprodpot}, using $b(\phi) = 0.1\phi$, we see that we still get much better results than the first order case. Making $\beta$ any larger than this, however, proves problematic as we quickly lose track of the amplitude of the power spectra whilst retaining the general form of the evolution. This is not, as one might initially assume, indicative of a problem with the second order terms that are the subject of this paper as we can remove these from equations (\ref{Geq}) and see that the erroneous amplitude is coming entirely from the first order terms. This is not unexpected as the smallness of each of the slow roll parameters for the majority of the inflationary period implies that they would have little effect at second order.  When using values of $\beta > 0.1$, $\phi$ can quickly get pulled away from its minimum to field values $\gg 10$ leaving our coupling term to grow unreasonably large --- $e^{\beta\phi} \gg 20000$ --- which in turn leads to a positive feedback loop. This just means that in cases such as these, more thought is required to result in reasonable, and viable inflationary trajectories. So whilst not correctly approximating the final expected amplitude of the power spectra, the features within the evolution can still be captured and can be better understood using this breakdown of slow roll terms than would otherwise be possible. In addition to this, it should be noted that in less extreme cases, where the inflaton is allowed to settle down to its minimum with less interruption from the coupling, the agreement is much better.

\subsection{$b(\phi) = \beta \phi^2$ example - double quadratic potential}

\begin{figure}
\includegraphics[width=\linewidth]{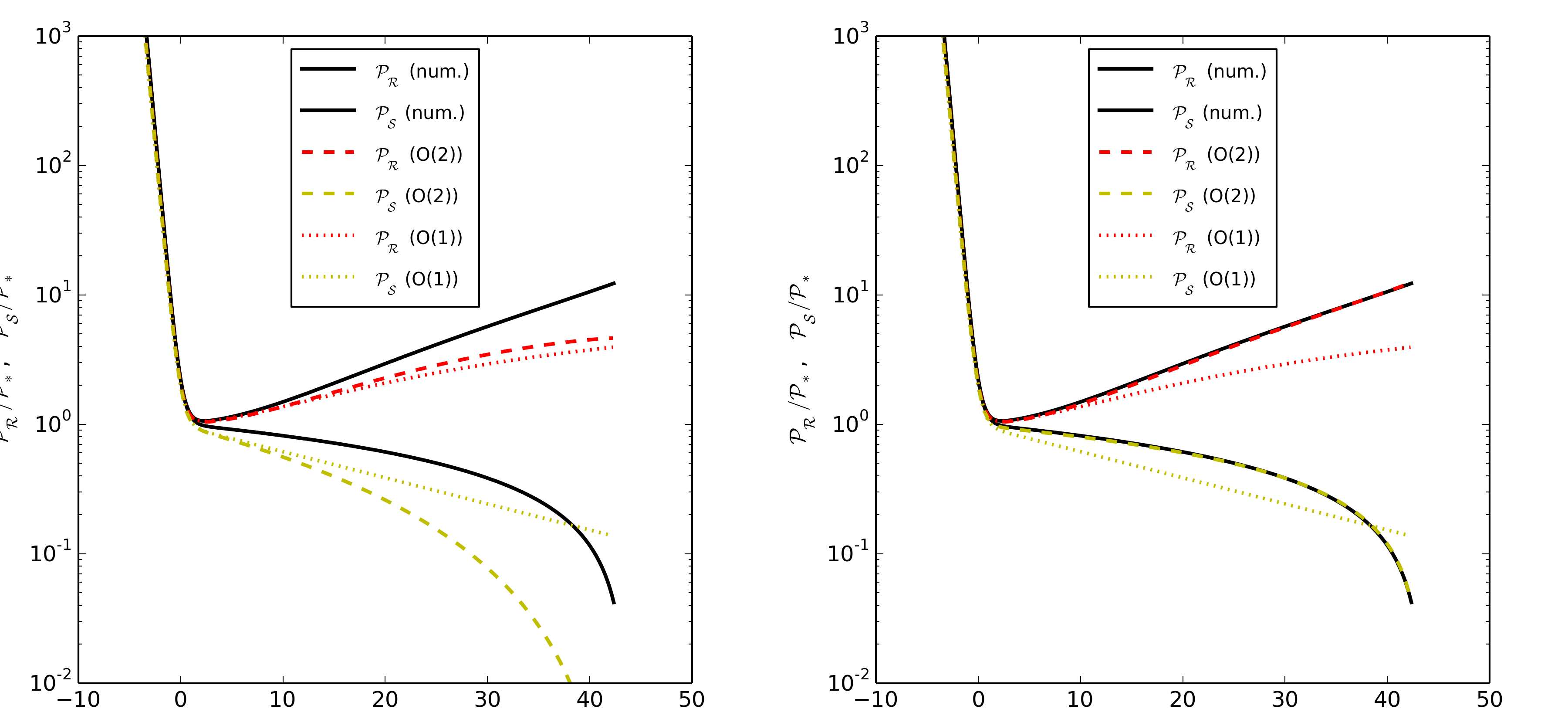}
  \caption{Double quadratic inflation with a higher order coupling term ($b(\phi) = \phi^2$). Here, the left graph power spectra with the solid black lines representing the numerical results, the dotted lines representing the first order results and the dashed lines representing the second order results --- \textbf{without} the new second order non-canonical terms ($\xi_2$) included. The right hand graph shows exactly the same, but with all terms included.}
  \label{phi2}
\end{figure}

Finally, a return to the double quadratic potential is in order to have a brief look at the effect of using higher powers of $\phi$ within $b(\phi)$. This is useful in order to look at the effect that $\xi_2$ has on the accuracy of our results --- as it has previously been neglected in our $b(\phi) = \beta\phi$ models. With a non-canonical term of this form, it was decided to use as similar a model as possible to that of the earlier non-canonical double quadratic case (with $b(\phi) = \phi$) --- but it was necessary to pull the initial $\phi$ value away from zero, whilst keeping it sufficiently small so as to avoid a runaway feedback loop (as mentioned in section 4.4), since using $\phi_0 = 0$ would have lead to an effectively single field case due to the presence of $\phi$ in the $b_\phi$ derivatives in the background equations.  Figure \ref{phi2} shows the power spectra of the same inflationary model but with two runs, the left where the $\xi_2$ terms have been left out and the right where they have been included. In the first case, it is clear that the analytical expressions fail to properly track the power spectra, but when the full expressions are used the agreement is perfect --- as the analytic results overlap the numerics throughout.

\section{Conclusion}
In this paper we have looked at the effect of non-canonical kinetic terms in the form of $e^{b(\phi)}$ where where we have set $b(\phi) = \beta\phi$ or $b(\phi) = \beta\phi^2$. The effects of such a term have important consequences for the shape of the background trajectory and hence the slow roll parameters --- of which we introduced a new parameter $\xi_2$ following on from the definition of $\xi_1$ in \cite{Lalak07}. Whilst the majority of cases studied involve only the linear non-canonical term and therefore leave $\xi_2 = 0$, the final example shows the importance of these corrections when higher powers of $\phi$ are used in the exponential --- providing the basis for further work with such conditions.  We have also seen that in some cases (eg. non-canonical double inflation) it is easily justified that second order $\xi_1$ terms are necessary to track the evolution accurately as they dominate over other slow roll parameters through both the sub-horizon and super-horizon regimes. Unfortunately, this generalisation of the canonical second order approach to include these non-canonical terms does lose a certain amount of accuracy when compared to the success of the canonical approach , but these differences are best explained not by errors in the expansion itself, but rather as problems in the altered field space trajectory. For example, the product potential (figure \ref{canprodpot}) behaves well with slow turns in the trajectory when $b(\phi) = 0$ but when $b(\phi) \not= 0$ (figure \ref{noncanprodpot}) we see a much more abrupt change of direction and subsequent loss of accuracy. The inclusion of a non-canonical term alters all of the slow roll parameters through this change of trajectory such that the errors are accounted for not only by $\xi_{1,2}$ but also by both the first and second order `normal' slow roll parameters --- of which we find agreement with earlier work.

We finally emphasise that whilst we do find an improvement on previous approximations through the inclusion of second order non-canonical terms it is still necessary to artificially constrain the amplitude of $b(\phi)$ in order for the approximation to work. This can be done either by setting $\phi_0$ or $\beta$ to be small --- as without doing so $\xi_{1,2}$ can become much larger than the other slow roll parameters much earlier on.

\acknowledgments
We are grateful to Anastasios Avgoustidis for helpful discussions concerning the numerical methods used. CvdB is supported by the Lancaster-Manchester-Sheffield Consortium for Fundamental Physics under STFC grant ST/J000418/1.

\appendix
\section{Appendix A: Second order expansions of slow-roll parameters}

It is relatively easy to show that the time dependence of $\epsilon$ goes as
\begin{eqnarray}
\dot{\epsilon} = 2H\epsilon(2\epsilon-\eta_{\sigma\sigma}-\xi_1 s^2_\theta c_\theta),
\end{eqnarray}
but a slightly less trivial task comes in the expansion of the $\eta_{IJ}$ parameters. In the following we will need the relation, to first order at least
\begin{eqnarray}\label{thetadot}
\dot{\theta} = -H\left(\eta_{\sigma s} + \xi_1s_\theta c_\theta^2\right),
\end{eqnarray}
in order to calculate some of the time derivatives. Now, starting from the definition of $\eta_{IJ}$ and differentiating it.
\begin{eqnarray}
\eta_{IJ} &=& \frac{V_{IJ}}{3H^2},\label{generaleta}\\
\dot{\eta_{ss}} &=& -\frac{2V_{ss}\dot{H}}{3H^3}+\frac{\dot{V_{ss}}}{3H^2},\label{etassdot}
\end{eqnarray}
where in this instance we have used I,J = s as an example.
The first term on the right hand side of Eq.(\ref{etassdot}) can easily be calculated from the definition of $\epsilon$ and $\eta_{IJ}$, however, for the second term we must use the definition of $V_{ss}$ to find $\dot{V_{ss}}$:
\begin{eqnarray}
V_{ss} = \sin^2\theta V_{\phi\phi} - 2\sin\theta \cos\theta e^{-b}V_{\phi\chi} + e^{-2b}\cos^2\theta V_{\chi\chi}.
\end{eqnarray}
After differentiating this and following some algebra, we arrive at
\begin{eqnarray}
\dot{V_{ss}} = 2\dot{\theta}(-V_{\sigma s})+\dot{\sigma}\left(V_{\sigma ss} + 2\cos^2(\theta)b_\phi e^{-b}(\sin\theta V_{\phi\chi}-\cos\theta e^{-b}V_{\chi\chi})\right).
\end{eqnarray}
Now, from Eq.(\ref{etassdot}) - using the definitions of the slow roll parameters and Eq.(\ref{thetadot}), we can show that 

\begin{align}
\dot{\eta_{ss}}&=2H\epsilon\eta_{ss}+\frac{1}{3H^2}\left[-2V_{\sigma s}\dot{\theta}+ \dot{\sigma}\left(2\cos^2(\theta)b_\phi e^{-b}(\sin\theta V_{\phi\chi}-\cos\theta e^{-b}V_{\chi\chi})\right)\right] - H\alpha_{\sigma ss}\\
&=2H\epsilon\eta_{ss}+2\eta_{\sigma s}(H\eta_{\sigma s}+H\xi_1s_\theta c_\theta^2)- 2Hc^2_\theta s_\theta \xi_1\eta_{\sigma s} -2Hc^3_\theta\xi_1\eta_{ss} - H\alpha_{\sigma ss}\\
&=2H\epsilon\eta_{ss}+2H\eta_{\sigma s}^2-2Hc^3_\theta\xi_1\eta_{ss} - H\alpha_{\sigma ss},
\end{align}
where we have defined:
\begin{equation}
\alpha_{IJK}\equiv \frac{V_\sigma V_{IJK}}{V^2}.
\end{equation}
Similarly, we can find
\begin{eqnarray}
\dot{\eta_{\sigma\sigma}} &=& 2H\epsilon\eta_{\sigma\sigma} - 2H\epsilon\eta_{\sigma s}-2H\eta_{\sigma\sigma}\xi_1 s^2_\theta c_\theta -4H\eta_{\sigma s}\xi_1 s_\theta c_\theta^2 - H\alpha_{\sigma\sigma\sigma},\\
\dot{\eta_{\sigma s}} &=& 2H\epsilon\eta_{\sigma s} +H\eta_{\sigma s}\eta_{\sigma \sigma}-H\eta_{\sigma s}\eta_{ss} -2H\eta_{ss}\xi_1 s_\theta c_\theta^2 -H\eta_{\sigma s}\xi_1 c_\theta - H\alpha_{\sigma\sigma s}.
\end{eqnarray}
The remaining first order slow-roll parameter, $\xi_1$, can be expanded as 
\begin{eqnarray}
\dot{\xi_1} &=& \frac{b_\phi \dot{\epsilon}}{\sqrt{2\epsilon}} + \sqrt{2\epsilon}b_{\phi\phi}\dot{\phi}\\
&=& \frac{4Hb_\phi \epsilon^2}{\sqrt{2\epsilon}}-\frac{\sqrt{2}Hb_\phi\epsilon\eta_{\sigma\sigma}}{\sqrt{\epsilon}}-\frac{\sqrt{2}Hb_\phi\epsilon\xi_1 s^2_\theta c_\theta}{\sqrt{\epsilon}}+ \sqrt{2\epsilon}b_{\phi\phi}\dot{\sigma}c_\theta\\
&=& 2H\epsilon\xi_1 - H\xi_1\eta_{\sigma\sigma}- H\xi_1^2 s^2_\theta c_\theta + 2H\epsilon b_{\phi\phi}c_\theta\\
&=& 2H\epsilon\xi_1 - H\xi_1\eta_{\sigma\sigma}- H\xi_1^2 s^2_\theta c_\theta + H\xi_2c_\theta \label{xi1dot}.
\end{eqnarray}

\end{document}